\documentclass{aa}
\usepackage{txfonts}
\usepackage{times}
\usepackage{graphicx}
\usepackage{natbib}
\usepackage{mathptm}

\begin{document}

\title{An ultra-cool white dwarf serendipitously found with COMBO-17}

\author{C. Wolf\inst{1} }
\offprints{C. Wolf, email: cwolf@astro.ox.ac.uk}
\institute{ Department of Physics, Denys Wilkinson Bldg.,
            University of Oxford, Keble Road, Oxford, OX1 3RH, U.K. }

\date{Received 12 October 2005 / Accepted }

\abstract{
We report the discovery of an ultra-cool white dwarf in the COMBO-17 survey. So 
far, only seven objects have been discovered in this rare category of white 
dwarfs, which are characterized by strong flux depression in the far-red and 
near-infrared part of the spectrum, presumably due to collisionally induced 
absorption (CIA). The new object COMBO-17 J114356.08-0144032 has very similar 
colours to LHS 3250, which was the first of its kind to be recognized. However, 
at $R=21.5$ it is the faintest and possibly most distant such object discovered
to date. It is the only such object in COMBO-17 at $R<23$; due to the small
sky coverage of 0.78~$\sq\degr$ this chance discovery can not provide any
constraints on the abundance of faint ultra-cool white dwarfs. We speculate on 
the basis of the proper motion that this new object is probably a member of 
the Galactic (thin) disk.
}
\titlerunning{An ultra-cool white dwarf in COMBO-17}
\authorrunning{Wolf}
\maketitle

\section{Introduction}

Cool white dwarfs in the disk of the Milky Way can reveal its age and give
clues about its early star formation history. Most of the uncertainty in 
the age is propagated from the cooling tracks of white dwarfs and not so
much given by observational samples of cool white dwarfs. However, it
came as a surprise that white dwarfs below 5000~K surface temperature do
not show redder colours with lower temperature. Instead, collisionally
induced absorption (CIA) by hydrogen molecules supresses the near-infrared 
spectrum \cite{Be94}.

Recently, a new terminology has differentiated between {\it cool} and {\it
ultra-cool} white dwarfs, which have temperatures above and below 4000~K,
respectively. In ultra-cool white dwarfs CIA is so strong that besides the
near-infrared also optical colours are affected, and the white dwarf colour 
tracks in optical colour diagrams are reverted. The stronger CIA pushes the 
peak of the spectrum further to the blue with decreasing temperature.

LHS 3250 was the first ultra-cool white dwarf recognized as such after its
parallax and hence luminosity were measured \cite{Ha99}. Six more objects 
have been found since, using SDSS $gri$ colours \cite{Ha01,Ga04} or large 
proper motion \cite{Op01}.
Currently outstanding questions are the accurate physical description of
these objects and their membership to the various Galactic components,
i.e. thin disk, thick disk and halo. Presently, the favoured scenario
assumes these objects to be old, cool white dwarfs with Helium cores and
very low masses of $\sim 1/4~M_{\odot}$ \cite{BL02}. Modeling the
spectra is still a challenge, but $\sim$3000~K Helium atmospheres with a 
$10^{-5}$ contribution from Hydrogen seem to be the most promising route.

This letter reports the serendipitous discovery of an ultra-cool white
dwarf in the COMBO-17 survey. COMBO-17 is a deep extragalactic multi-band
survey characterized by a range of medium-band filters facilitating {\it 
fuzzy spectroscopy}. The spectral resolution of the filters is sufficient 
to differentiate between regular stars, white dwarfs, galaxies and QSOs at 
$R\la 23$ and to obtain very accurate photometric redshifts. The new object 
is currently the faintest, and potentially most distant, known member of 
its class. All COMBO-17 photometry is reported as Vega-normalised 
magnitudes, while SDSS colours are given in the AB system.

\section{Data}

\subsection{COMBO-17 photometry and selection}

\begin{figure}
\centering
\includegraphics[clip,angle=270,width=0.85\hsize]{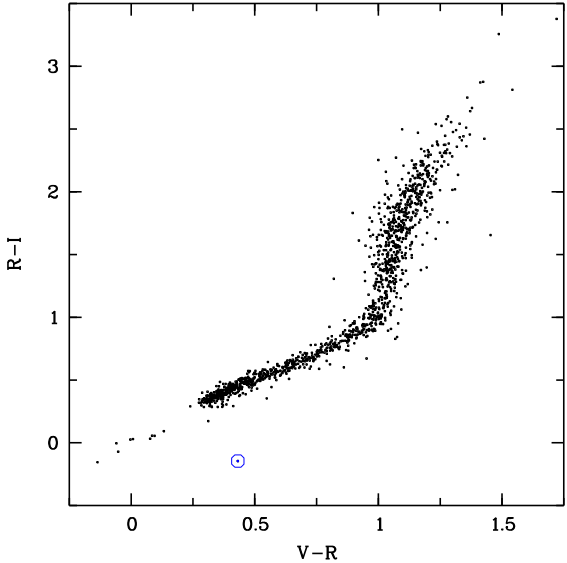}
\caption{Colour-colour-diagram of COMBO-17-classified stars:
1332 stars and 9 white dwarfs with $T>10000$~K have been classified in the 
S11 field in the range of $R=[16,23]$. The object COMBO-17 J1143 marked here 
by a circle displays large template fitting residuals that are also reflected
by its offset from the stellar locus. 
\label{CCD}}
\end{figure}

An unusual object was serendipitously discovered during an inspection of 
colour-colour diagrams of the S11 field in COMBO-17. The respective diagram 
showed a sample of objects at $R=[16,23]$ which were classified by their SED
as stars or white dwarfs (see Wolf et al. 2004 for a discussion of the SED 
classification in COMBO-17). One object, COMBO-17 J114356.08-0144032 
(hereafter COMBO-17 J1143), was noted in a location clearly inconsistent with 
the stellar colours (see Fig.~\ref{CCD}). It is isolated on the sky and not 
affected by bright neighbors. It shows no signs of magnitude variability above 
5\% on time scales from weeks to a few years and it is morphologically unresolved.

Evidently, all stars in the stellar template library are inconsistent with 
its SED, even when considering composite spectra of binaries. Also, all galaxy 
and AGN templates are even less consistent with the SED. Hence, we have to 
suspect that this object is of a more unusual kind.

We searched the three COMBO-17 fields for objects with similar colours, but
at $R<23$ we only find objects which are clearly QSOs. Fainter than $R=23$
our photometry is too noisy to select such objects and differentiate them
from QSOs. Hence, our survey contains exactly one object of this kind in an
area of 0.78~$\sq\degr$ at $R<23$.

With $R\sim 21.5$ the object is two magnitudes fainter than the faintest 
other known ultra-cool white dwarf SDSS J1001. It is well detected in all 
17 passbands (see Table~1). A J-band upper limit is available from the
{\it COMBO-17+4} observations that extend these SEDs into the near-infrared
domain (PI Meisenheimer). A 3$\sigma$ upper limit of $J<21.6$ is obtained
from aperture photometry on the known position of the object. Formally, the 
object is measured with $J=22.8$ and a signal-to-noise ratio around 1.

\begin{figure}
\centering
\includegraphics[clip,angle=270,width=0.9\hsize]{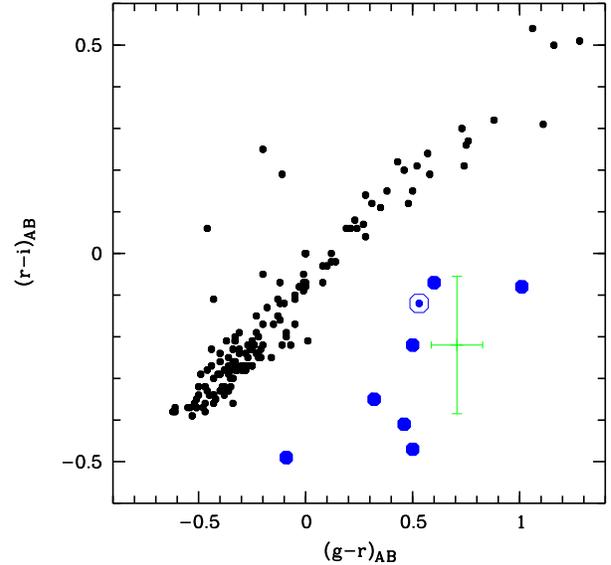}
\caption{Colour-colour-diagram of white dwarfs:
Large symbols show all known ultra-cool white dwarfs, where COMBO-17 J1143
is marked by a circle. Its original SDSS photometry is shown as the data 
point with error bars. Normal white dwarfs are shown as small symbols (sample 
by Kleinman et al. 2004). Objects which are redder than normal white dwarfs 
in $r-i$ have M dwarf companions while bluer ones show CIA in optical bands.
\label{CWDs}}
\end{figure}

We have also found the object in the SDSS database, where it is listed as
a faint QSO candidate due to its location in the gri colour diagram. From
SDSS data alone it could not have been identified as an ultra-cool white 
dwarf candidate, because the z band is too noisy to differentiate it
against QSOs, which are vastly more common at this magnitude. The SDSS 
colour indices (corrected for negligible interstellar reddening) are $g-r=
0.71\pm 0.12$ and $r-i=-0.23 \pm 0.17$. The deeper data from COMBO-17 allow 
to estimate more accurate SDSS colours by comparing the synthetic COMBO-17 
colours of the seven known objects as calculated from their spectra. Thus,
we find $g-r=0.53\pm 0.05$ and $r-i=-0.12\pm 0.05$.

\subsection{Astrometry and proper motion}

Our observations of the S11 field include four imaging epochs from February 
1999 to May 2002, which we can compare astrometrically to search for proper 
motion. The COMBO-17 data reduction has been optimized for photometry but not 
for accurate astrometry, and general astrometry has internal errors of $\sim 
0.15\arcsec$ in every epoch. We achieved the best local astrometric accuracy 
by using only reference objects within $2\arcmin$ of the target object, while
including also galaxies to provide the reference system. Some stars have been 
excluded from the astrometric fit because of their own proper motion. We only 
compare images taken through identical filters to avoid colour-dependent 
displacement of individual objects as a result of differential refraction. 
This leaves us with B-band imaging in February 1999 and January 2001 (PSF 
on co-added frame $1\farcs 1$) as well as R-band imaging in January 2000 
and May 2002 (PSF $0\farcs 8$). 

The positional scatter among 35 galaxies suggests an rms error in the measured 
position difference of $0\farcs 026$ and $0\farcs 017$ in the B-band and R-band comparisons, 
separated by 1.94 yr and 2.21 yr, respectively. The R-band comparison could 
be biased by parallax if the object was closer than 50 pc. We measure strong 
proper motion signals of $0\farcs 046$/yr for a nearby M3 dwarf with $R=19$
and of $\sim 0\farcs 052$/yr for the white dwarf COMBO-17 J1143, each with an 
error of $\sim 0\farcs 010$/yr. The M dwarf is moving in an almost opposite
direction, and there is no proper motion companion to the white dwarf among 
the reference objects.

\section{Discussion}

\begin{figure}
\centering
\includegraphics[clip,angle=270,width=0.95\hsize]{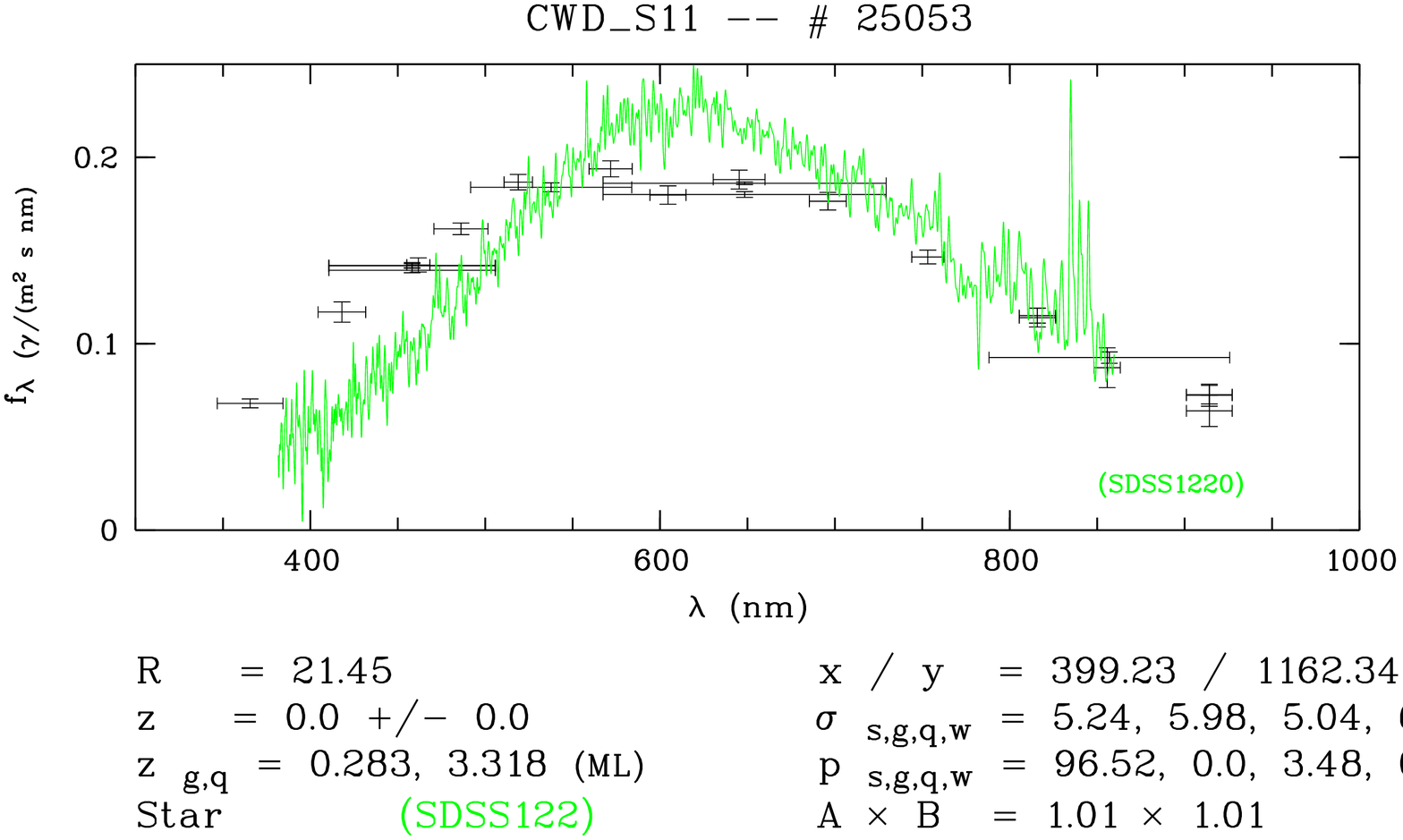}
\includegraphics[clip,angle=270,width=0.95\hsize]{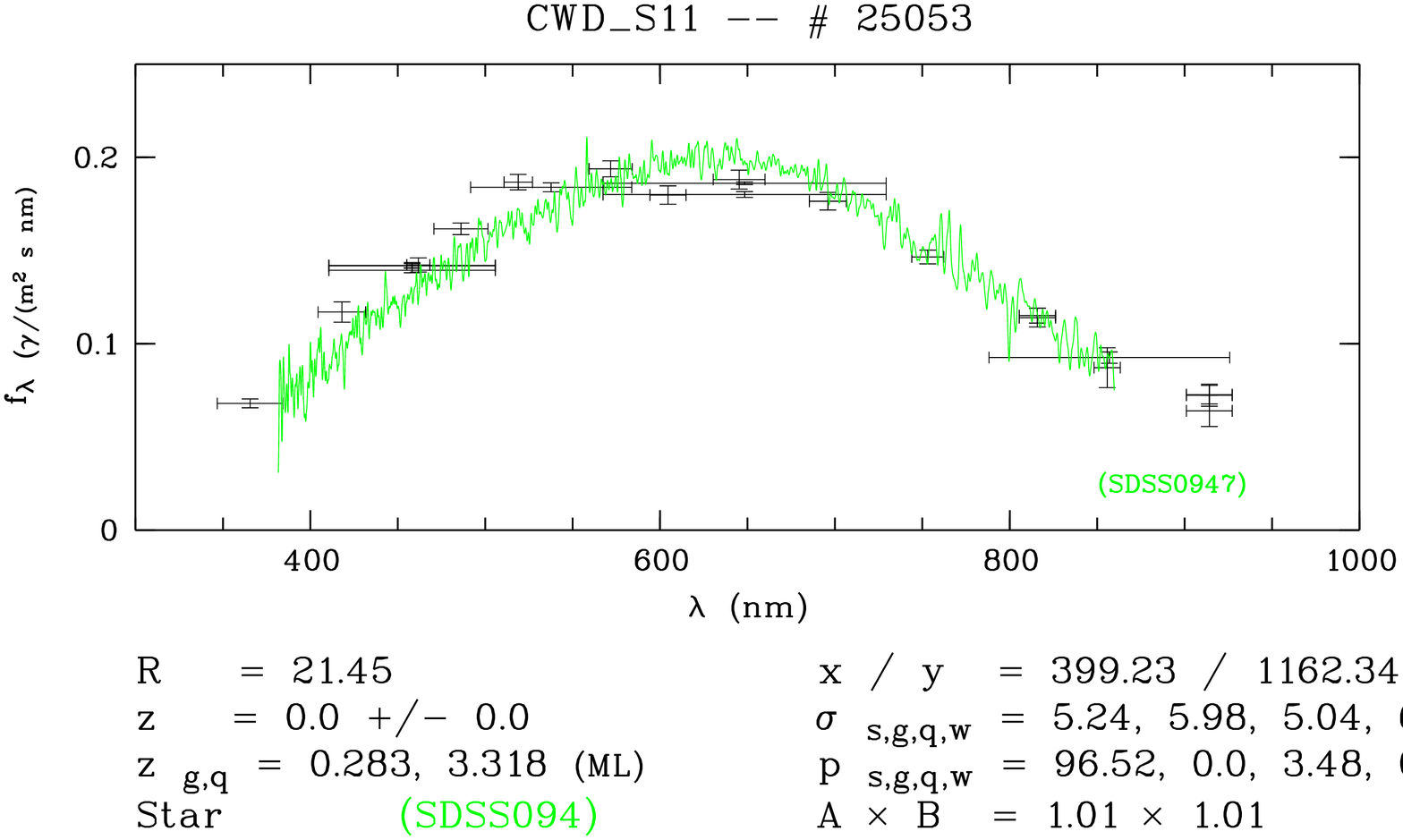}
\includegraphics[clip,angle=270,width=0.95\hsize]{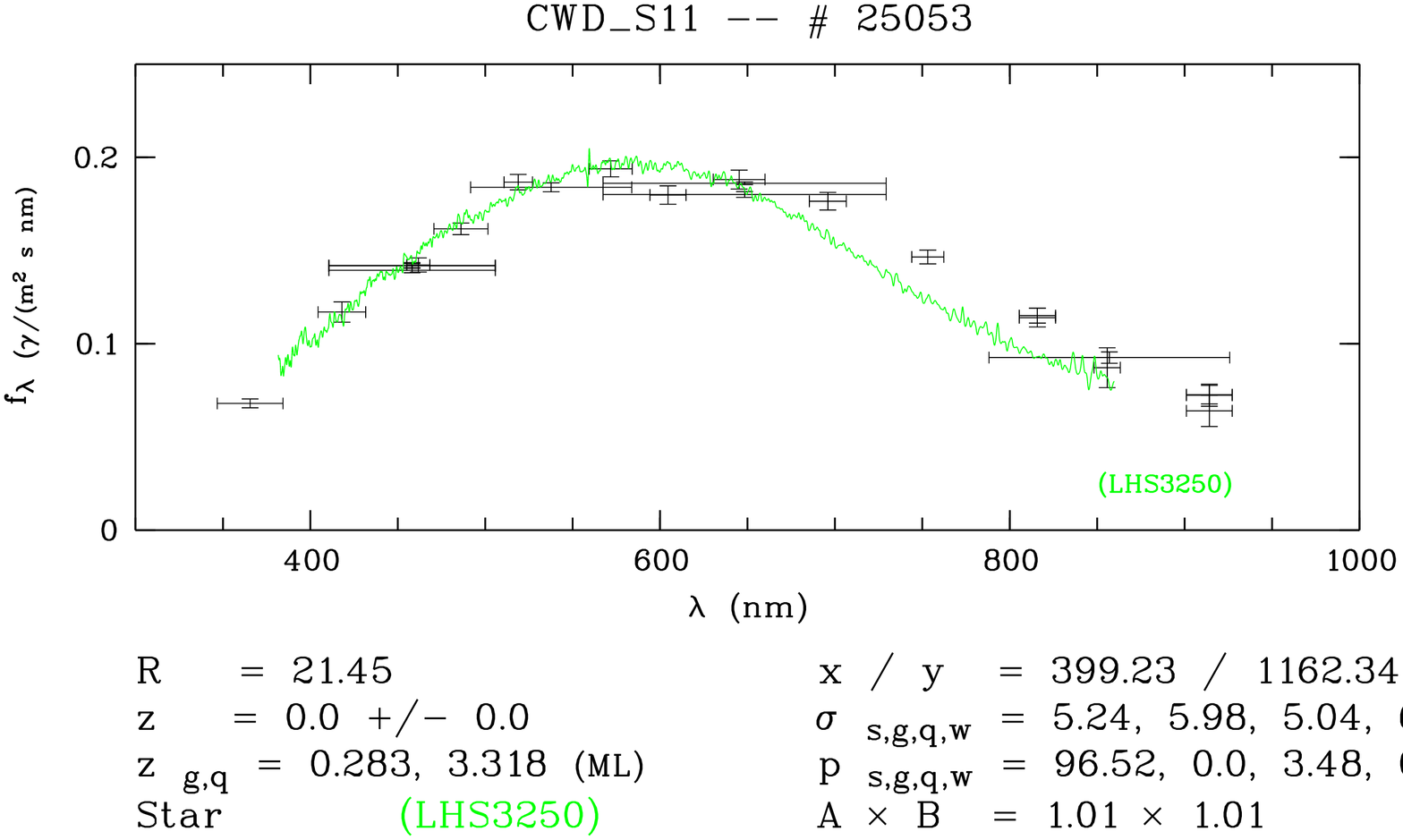}
\includegraphics[clip,angle=270,width=0.95\hsize]{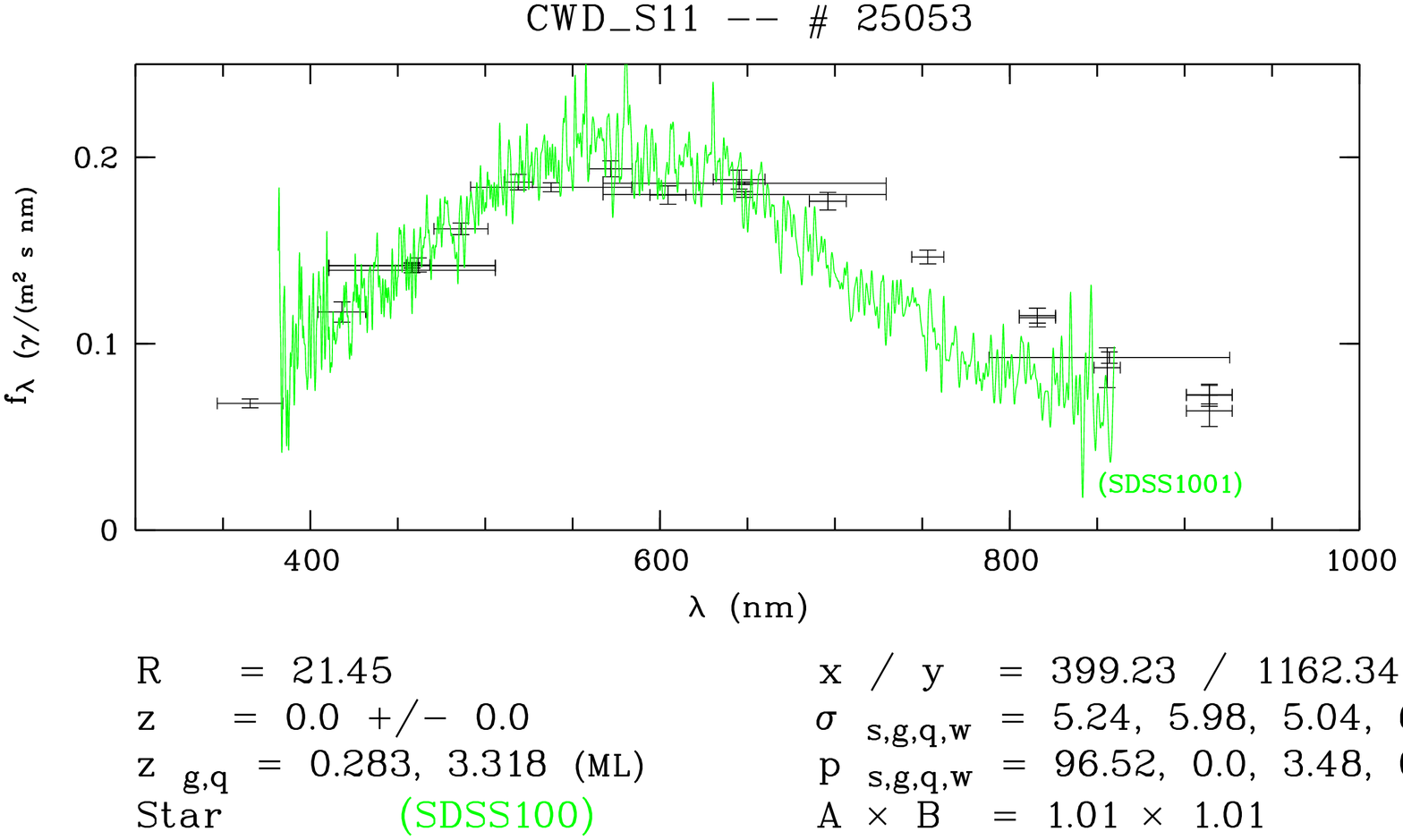}
\includegraphics[clip,angle=270,width=0.95\hsize]{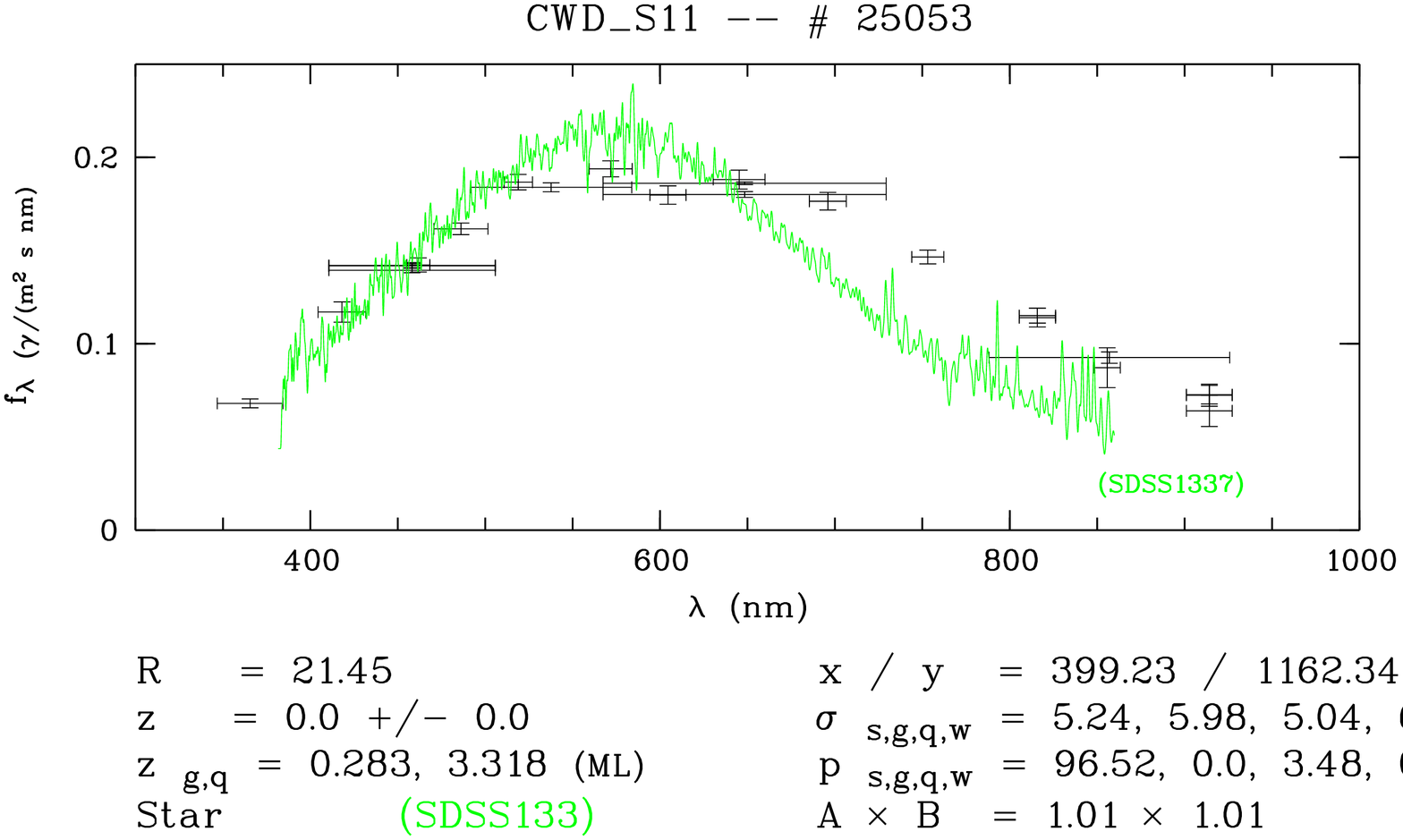}
\includegraphics[clip,angle=270,width=0.95\hsize]{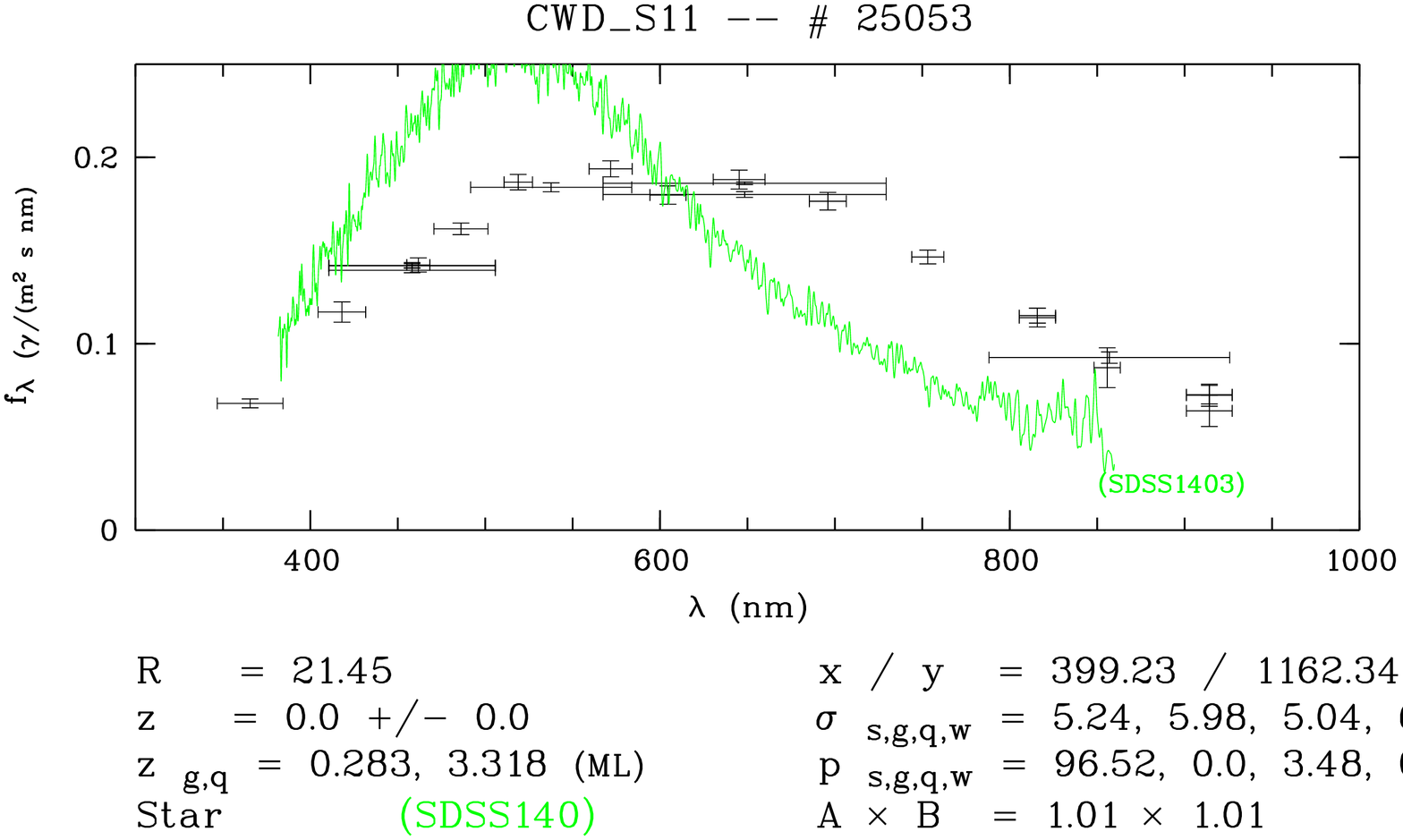}
\caption{The SED of COMBO-17-J1143 (black data points) compared to the spectra 
of all known ultra-cool white dwarfs (grey lines) except for LHS 1402. COMBO-17
photometry is shown with horizontal error bars indicating filter width and 
vertical error bars representing the respective flux error. The spectra have 
been normalised with respect to the measured flux in the three filters B, V and 
R. COMBO-17 J1143 is most similar to LHS 3250 and SDSS J0947.
\label{spec}}
\end{figure}

\begin{figure}
\centering
\includegraphics[clip,angle=0,width=0.8\hsize]{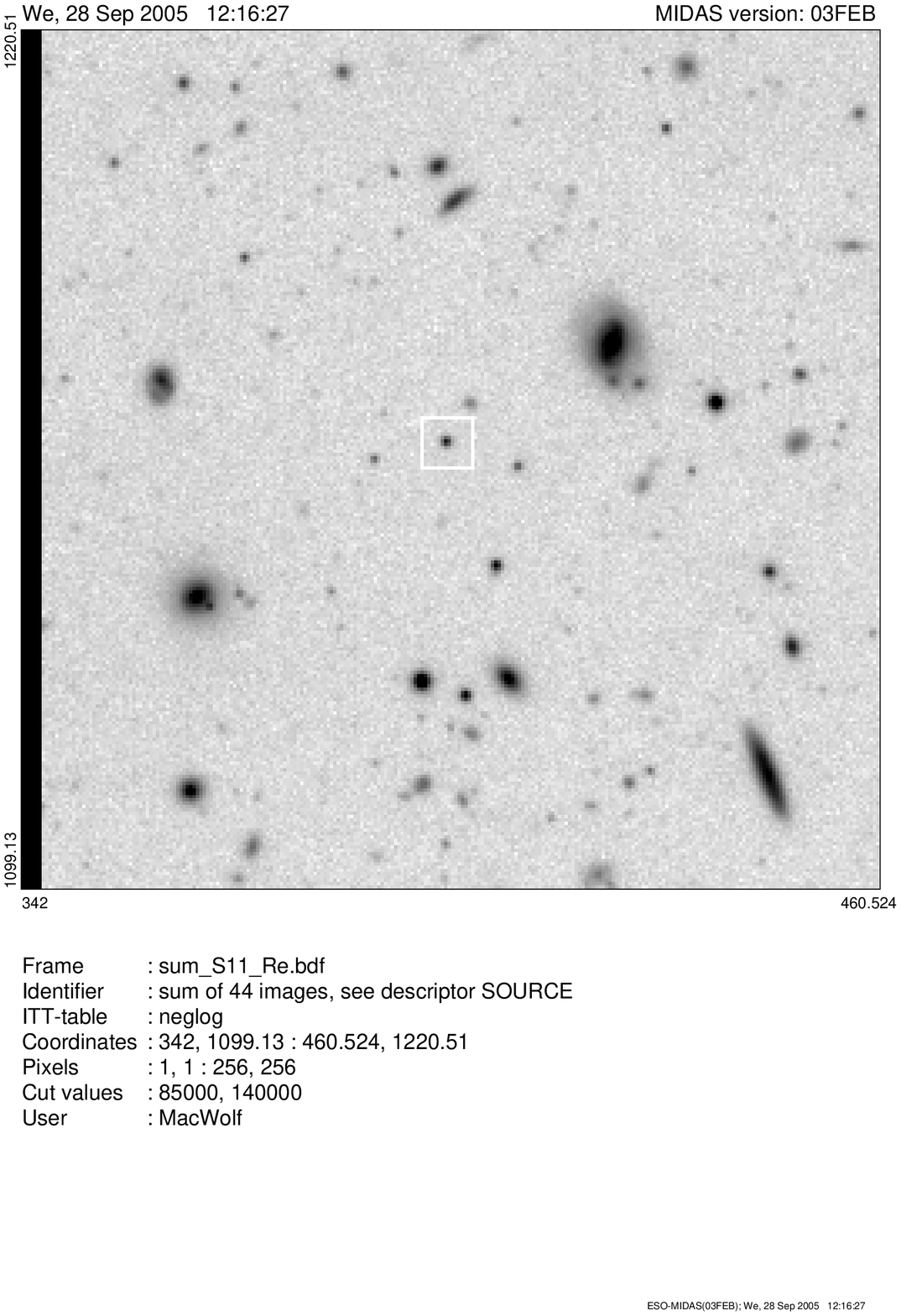}
\caption{Finding chart for COMBO-17 J114356.08-0144032 taken from a 5-hour 
R-band image obtained with WFI at the 2.2~m-telescope at La Silla in February 
2000. The frame is $120\arcsec$ on a side. North is up and East is on the left.
\label{fc}}
\end{figure}

\subsection{Disk or Halo object?}

Assuming an uncertain luminosity of $M_V=16.5 \pm 1.0$ for objects of this 
kind and colour (following Salim et al. 2004) places COMBO-17-J1143 at a 
distance of 118~pc with 0.2~dex error ($^{+70}_{-45}$~pc). This translates 
into a tangential velocity of $v_{\rm tan}=29$~km/s relative to the Sun.
$UVW$ velocities after transformation into the local standard of rest (LSR),
which is given as $(U,V,W)_\odot =(10.0,5.2,7.2)$~km/s by Binney \& Merrifield 
(1998), are listed in Tab.~1, whereby the radial velocity remains unknown. 
We compare these numbers with the velocity ellipsoids from Binney 
\& Merrifield (1998) for the Galactic components:

\noindent {\it Thin disk:}
$(\sigma_U,\sigma_V,\sigma_W)=(34,21,18)$~km/s, $\langle V 
\rangle = -6$~km/s.

\noindent {\it Thick disk:}
$(\sigma_U,\sigma_V,\sigma_W)=(61,58,39)$~km/s, $\langle V 
\rangle = -36$~km/s.

\noindent {\it Halo:}
$(\sigma_U,\sigma_V,\sigma_W)=(135,105,90)$~km/s, $\langle V 
\rangle = -185$~km/s.

Hence, we determine the likelihood ratio for membership to these three 
components based only on the proper motion and the ellipsoids but not 
considering the errors on the proper motion, whereby we marginalise over
the unknown radial velocity within $v_{\rm rad}=[-300,300]$~km/s. We get
$(p_{\rm thin},p_{\rm thick},p_{\rm halo}) = (0.82,0.16,0.02)$. Membership 
of the thin disk appears to be four times more likely than of the thick 
disk. Furthermore, the local density normalisation of cool white dwarfs 
and hence their a-priori chance of having them in a volume-limited sample
is $(q_{\rm thin},q_{\rm thick},q_{\rm halo}) = (0.69,0.24,0.07)$ (from  
Napiwotzki, private communication, based on Pauli et al., 2003). If cool 
white dwarfs are any guide to the currently unknown normalisations of 
ultra-cool objects, then thin disk membership would be more than 90\% 
likely on the whole. We note, that thin and thick disk membership would 
only be equally probable, if the distance was larger by a factor 
of $\sim 2.5$. Also, Gates et al. (2004) ruled out halo membership for 
most of their objects.

\subsection{Spectral features}

Unfortunately, the featureless spectra of ultra-cool dwarfs hold no prospect 
for measuring radial velocities or investigate potential binarity. If the 
object was a double degenerate as it is assumed for LHS 3250 because of its
luminosity of $M_V=15.7$, this would affect the distance measurement.

The filter
SED may suggest a somwewhat surprising dent around 600~nm, but that may be a 
chance outlier. Among 17 filters, one filter measurement should be expected 
to deviate by $\sim 1.8 \sigma$ from the true flux purely by chance. Another
similar dent in the spectrum of SDSS J1001 might not be real either given 
that the spectrum is rather noisy.

\subsection{Surface densities}

The recent SDSS search by Gates et al. (2004) found six objects in 4330~$\sq
\degr$, i.e. a surface density of $0.0014/\sq\degr$ with $i<20.2$, which is 
roughly equivalent to $R<19.8$. Given our $\sim 3$~mag deeper selection, we 
still expect to find only one ultra-cool dwarf within roughly a couple 
hundred square degrees. Our small survey area of 0.78~$\sq\degr$ demonstrates 
that COMBO-17 contains an ultra-cool white dwarf at $R<23$ purely by chance. 

This letter has reported another ultra-cool white dwarf, enlarging the sample
of known cases to eight. Although, it is presumably the most distant example
of its kind, it appears not particularly likely to belong to the Galactic 
thick disk or halo. We will need measurements of parallaxes to constrain the
physical nature of these objects and proceed further to an understanding of
their atmospheres. If spectral features were found and radial velocities and 
distances were measured, the kinematic properties of ultra-cool white 
dwarfs as a group could reveal clues about the stellar evolution channel
responsible for their origin.

\begin{table}
\caption{Astrometry of the white dwarf COMBO-17 J1143.
\label{objectast}}
\begin{tabular}{lr}
\hline \noalign{\smallskip} \hline \noalign{\smallskip} 
$\alpha$ (J2000) run E	&  $11^h 43^m 56\fs08$  \\
$\delta$ (J2000) run E  &  $-01\degr 44\arcmin 03\farcs 21$  \\  
$l_{\rm gal}$		&  $270\fdg 944$ \\
$b_{\rm gal}$		&  $56\fdg 6172$ \\
\noalign{\smallskip} \hline \noalign{\smallskip}
$\Delta \alpha_{\rm G-A}$	& $-0\farcs 110 \pm 0\farcs 025$ \\
$\Delta \delta_{\rm G-A}$	& $-0\farcs 039 \pm 0\farcs 028$ \\
$\Delta \alpha_{\rm H-E}$	& $-0\farcs 105 \pm 0\farcs 016$ \\
$\Delta \delta_{\rm H-E}$	& $-0\farcs 026 \pm 0\farcs 018$ \\
$\Delta \alpha_{\rm tot}$/yr	& $-0\farcs 050 \pm 0\farcs 006$ \\
$\Delta \delta_{\rm tot}$/yr	& $-0\farcs 014 \pm 0\farcs 007$ \\
\noalign{\smallskip} \hline \noalign{\smallskip}
Julian epoch B-band run A	& 1999.122 \\
Julian epoch R-band run E	& 2000.112 \\
Julian epoch B-band run G	& 2001.063 \\
Julian epoch R-band run H	& 2002.332 \\
\noalign{\smallskip} \hline \noalign{\smallskip}
$v_{\rm tan}$	&  29~km/s \\
$v_{\rm rad}$	&  unknown \\
$U_{\rm LSR}$	&  (-10.8-0.009~$v_{\rm rad}$)~km/s \\
$V_{\rm LSR}$	&  (-11.8-0.550~$v_{\rm rad}$)~km/s \\
$W_{\rm LSR}$	&  ( -3.8+0.835~$v_{\rm rad}$)~km/s \\
\noalign{\smallskip} \hline
\end{tabular}
\end{table}

\begin{table}
\caption{Photometry of the white dwarf COMBO-17 J1143. A minimum photometric 
error of $0\fm 03$ was assumed to take calibration uncertainties into account.
\label{objectphot}}
\begin{tabular}{llcc}
\hline \noalign{\smallskip} \hline \noalign{\smallskip}
\multicolumn{2}{c}{$\lambda_\mathrm{\mathrm{cen}}$/fwhm}  & Vega-mag & ABmag \\
\noalign{\smallskip} \hline \noalign{\smallskip} 
365/36  &$U$ \,\,\, 	& $22.65 \pm 0.04$	& 23.44 \\
458/97  &$B$	& $22.45 \pm 0.03$	& 22.34 \\
538/89  &$V$ 	& $21.86 \pm 0.03$	& 21.86 \\
648/160 &$R$ 	& $21.46 \pm 0.03$	& 21.67 \\
857/147 &$I$	& $21.60 \pm 0.04$	& 22.11 \\
	&$J$	& $<21.6$		& $<22.57$ \\
418/27 & 	& $22.70 \pm 0.05$	& 22.54 \\
462/13 & 	& $22.45 \pm 0.03$	& 22.29 \\
486/31 & 	& $22.15 \pm 0.03$	& 22.11 \\
519/16 & 	& $21.93 \pm 0.03$	& 21.89 \\
572/25 & 	& $21.71 \pm 0.03$	& 21.77 \\
605/21 & 	& $21.66 \pm 0.03$	& 21.79 \\
645/30 & 	& $21.45 \pm 0.03$	& 21.69 \\
696/21 & 	& $21.36 \pm 0.03$	& 21.65 \\
753/18 & 	& $21.38 \pm 0.03$	& 21.77 \\
816/21 & 	& $21.47 \pm 0.04$	& 21.96 \\
857/15 & 	& $21.66 \pm 0.13$	& 22.24 \\
914/26 & 	& $21.83 \pm 0.06$	& 22.35 \\
\noalign{\smallskip} \hline \noalign{\smallskip}
$(g-r)_{\rm SDSS}$ & 	&	& $ 0.71 \pm 0.12$ \\
$(r-i)_{\rm SDSS}$ &	&	& $-0.23 \pm 0.17$ \\
$(g-r)_{\rm COMBO-17}$ & & 	& $ 0.53 \pm 0.05$ \\
$(r-i)_{\rm COMBO-17}$ & & 	& $-0.12 \pm 0.05$ \\
\noalign{\smallskip} \hline
\end{tabular}
\end{table}

\begin{acknowledgements}
Ralf Napiwotzki, Philipp Podsiadlowski, James Binney, Stephen Justham, 
Tony Lynas-Gray and Klaus Meisenheimer, as well as an anonymous referee 
have helped to improve the paper
with comments and discussion. Evalyn Gates has provided colours of 
previously known white dwarfs and Geza Gyuk the spectra of previously 
known ultra-cool white dwarfs. C.W. was supported by a PPARC Advanced 
Fellowship. Two colours of COMBO-17 J1143 were extracted from the 
SDSS Archive. 
\end{acknowledgements}

\end{document}